\documentstyle[preprint,aps]{revtex}
\begin{document}
\draft
\preprint{}

\title {\bf Factorization in processes of graviton \\
            scattering off electron for $Z$ and $W$ productions}
\vskip 0.5cm

\author{J. S. Shim}
\address{Department of Physics, Kyung Hee University,
         Seoul 130-742, Korea}

\author{H. S. Song}
\address{Center for Theoretical Physics and Department of
         Physics,\\
         Seoul National University, Seoul 151-742, Korea}

\maketitle

\begin{abstract}
The study of factorization in linearized gravity is extended to
the graviton scattering processes with an electron for the
massive vector boson productions such as $g e \rightarrow Z e$
and $g e \rightarrow  W \nu_e$.
It is shown that every transition amplitude is completely
factorized due to gravitational gauge invariance and Lorentz
invariance. Also the explicit values of vector boson polarizations
are obtained.

\vskip 0.4cm
\noindent
PACS number(s) : 04.60.+n, 12.25.+e, 13.88.+e
\end{abstract}



\vskip 3cm

In the standard gauge theory every four-body Born amplitude
with a massless gauge boson as an external particle has been
well known to be factorizable
\cite{Grose,Goebel,Dongpei,Brodsky}
into one factor which depends only on charge or other
internal-symmetry indices and the other
factor which depends on spin or polarization indices.

In our previous works \cite{Choi1,Choi2},
we have shown that gravitational gauge invariance and graviton
transversality force the transition amplitudes
of four-body graviton interactions to be factorized.
In particular, we have investigated in a more extensive way the
four-body graviton interactions like
$g(k_1) + X(p_1) \rightarrow \gamma(k_2)+X(p_2)$ and
$g(k_1)+X(p_1) \rightarrow g(k_2)+X(p_2)$
in the context of linearized gravity where $X$ is any kind of
particles with spin less than 2 or graviton itself.
We have found that every amplitude can be completely
factorized into a simple form
composed of a kinematic factor, QED-like Compton scattering form,
and another gauge invariant terms.
The factorization property can
be used as a powerful tool to investigate the gravitational
interactions and the polarization effects.

So far, the processes of a graviton conversion into a photon and
gravitational Compton scattering are considered.
Also we have considered the case that the incoming matter is the
same as the outgoing one in each process.
But in this paper, we extend the study of factorization in the
process of high energy graviton conversion into a massive vector
boson
such as  $ge \rightarrow Z e$, $ge \rightarrow W \nu_e$.
Also it is interesting to note that factorization is held even if
matter particles in the initial state and final state have
different masses as in the process $g e \rightarrow W \nu_e$.

The Lagrangian ${\cal L}_I$ of the interaction of an electron and
massive vector boson ($Z$ or $W$) with gravitational field is
given by
\begin{eqnarray}
{\cal L}_I&=&{\cal L}_{ge}(A)+{\cal L}_{gW}(A)+{\cal L}_{gZ}
             +{\cal L}_{geW} + {\cal L}_{geZ}
 \label{act},\\
{\cal L}_{ge}(A)&=&\sqrt{-g}\left[\frac{i}{2}\left(\bar{\psi}
                \gamma^\mu (\vec{\nabla}_\mu -ieA_\mu) \psi
              - \bar{\psi}(\stackrel{\leftarrow}{\nabla}_\mu
               +ieA_\mu) \gamma^\mu \psi\right)
                - m_e \bar{\psi}\psi \right] \label{lea},\\
{\cal L}_{gW}(A)
	    &=&-\frac{1}{2} \sqrt{-g}g^{\mu\nu}g^{\alpha\beta}
                   (D_\mu W_\alpha -D_\alpha W_\mu)^*
                   (D_\nu W_\beta -D_\beta W_\nu) \nonumber\\
                &&+\sqrt{-g}g^{\mu\nu} m_W^2 W_\mu ^* W_\nu
                 - ie \sqrt{-g}g^{\mu\nu}g^{\alpha\beta}
                   W_\mu^* W_\alpha F_{\nu\beta},\\
{\cal L}_{gZ}&=&-\frac{1}{4} \sqrt{-g}g^{\mu\nu}g^{\alpha\beta}
                   (\partial_\mu Z_\alpha -\partial_\alpha Z_\mu)
                   (\partial_\nu Z_\beta -\partial_\beta Z_\nu)
		   \nonumber\\
             &&+\frac{1}{2}\sqrt{-g}g^{\mu\nu} m_Z^2 Z_\mu  Z_\nu
                  ,\\
{\cal L}_{geW}&=&g_W \sqrt{-g}g^{\mu\nu}
                 \left[\bar{\psi_e}W^*_\mu
                \gamma_\nu (1-\gamma_5) \psi_{\nu_e}
            +{\bar{\psi}}_{\nu_e} W_\mu \gamma_\nu(1-\gamma_5)
               \psi_e \right] \label{geW},\\
{\cal L}_{geZ}&=&g_Z \sqrt{-g}g^{\mu\nu}
                 \bar{\psi}\gamma_\mu
             \left[\epsilon_L(1-\gamma_5)+\epsilon_R(1+\gamma_5)
		 \right] \psi Z_\nu
                 \label{geZ},
\end{eqnarray}
where $g, g_W, g_Z, \epsilon_L$, and $\epsilon_R$ are defined as
\begin{eqnarray}
&&g=det g_{\mu\nu},
\nonumber\\
&&g_W
    =\left(2^{-\frac{1}{2}} G_F m_W^2 \right)^{\frac{1}{2}},
 \  \ g_Z
    =\left(2^{\frac{1}{2}} G_F m_Z^2 \right)^{\frac{1}{2}},
 \nonumber\\
&&\epsilon_L=-\frac{1}{2}+\sin^2\theta_W,
 \  \  \epsilon_R=\sin^2\theta_W,
\end{eqnarray}
where $\theta_W$ is the Weinberg angle.

In the procedure one introduces a symmetric tensor field
$h_{\mu\nu}$ denoting the deviation of the metric tensor
$g_{\mu\nu}$ from the flat space Minkowski metric tensor
$\eta_{\mu\nu}$;
\begin{eqnarray}
g_{\mu\nu}=\eta_{\mu\nu}+\kappa h_{\mu\nu},
\end{eqnarray}
where $\kappa=\sqrt{32\pi G_N}$.
After the expansion any curved space geometrical object is expressed
as an infinite series in terms of $h_{\mu\nu}$.
The process of graviton scattering off electron for the $Z$ boson
production $ge \rightarrow Ze$ is made up of four Feynman diagrams
(See Fig.~1) and the explicit form of its tree-level amplitude
${\cal M}_{ge \rightarrow Ze}$ is determined from the Lagrangian
${\cal L}_{ge}$, ${\cal L}_{gZ}$, and ${\cal L}_{geZ}$ as
\begin{eqnarray}
{\cal M}_{ge\rightarrow Ze}
&=&{\cal M}^{Ze}_a +{\cal M}^{Ze}_b
  +{\cal M}^{Ze}_c +{\cal M}^{Ze}_d,
\end{eqnarray}
\begin{eqnarray}
&&{\cal M}^{Ze}_a=\frac{i\kappa g_Z}{4}
     \frac{(\epsilon_1\cdot p_1)}{(p_1\cdot k_1)}
   \bar{u}(p_2,s_2)\not\!{\epsilon^*_2}\left[
   \epsilon_L(1-\gamma_5) +\epsilon_R(1+\gamma_5)\right]
     (\not\!{k_1}\not\!{\epsilon_1}+2\epsilon_1\cdot p_1)
     u(p_1,s_1),
     \label{mz1}\\
&&{\cal M}^{Ze}_b
     =-\frac{i\kappa g_Z}{4}
     \frac{(\epsilon_1\cdot p_2)} {(p_2\cdot k_1)}
   \bar{u}(p_2,s_2)
     ( 2\epsilon_1\cdot p_2
     -\not\!{\epsilon_1}\not\!{k_1} )
     \!\not\!{\epsilon^*_2}\left[
   \epsilon_L(1-\gamma_5) +\epsilon_R(1+\gamma_5)\right]
      u(p_1,s_1),
      \label{mz2}\\
&&{\cal M}^{Ze}_c
     =\frac{i\kappa g_Z}{4}
      \frac{1}{(k_1\cdot k_2)}
      \bar{u}(p_2,s_2)\left[2(\epsilon_1\cdot k_2)
         \left\{(\epsilon_1\cdot\epsilon^*_2) \not\!{k_1}
         -(\epsilon^*_2\cdot k_1)\not\!{\epsilon_1}
      -(\epsilon_1\cdot k_2)\not\!{\epsilon^*_2}
       \right\} \right.\nonumber\\
&& \hskip 5cm
    \left.+2(k_1\cdot k_2) (\epsilon_1\cdot\epsilon^*_2)
       \not\!{\epsilon_1}\right] \left[
   \epsilon_L(1-\gamma_5) +\epsilon_R(1+\gamma_5)\right]
        u(p_1,s_1),
	\label{mz3}\\
&&{\cal M}^{Ze}_d
     =-\frac{i\kappa g_Z}{2}
 (\epsilon_1\cdot\epsilon^*_2)\bar{u}(p_2,s_2) \not\!{\epsilon_1}
   \left[\epsilon_L(1-\gamma_5) +\epsilon_R(1+\gamma_5)\right]
    u(p_1,s_1)
    \label{mz4},
\end{eqnarray}
where $\epsilon_1^\mu \epsilon_1^\nu$ and $k_1^\mu$ are the
polarization and momentum of the graviton, $\epsilon^{*\mu}_2$ and
$k_2^\mu$ are the polarization and momentum of the $Z$ boson
, and $p_1^\mu$ and $p_2^\mu$ are the initial and final momenta of
the massive electron.
{}From the above results, we can obtain the completely factorized
transition amplitude ${\cal M}_{ge\rightarrow Ze}$ as
\begin{eqnarray}
{\cal M}_{ge\rightarrow Ze}
&=& -\frac{\kappa}{2e}
     \frac{p_1\cdot k_1 p_2\cdot k_1}{k_1\cdot k_2}
     \left[\frac{p_1\cdot \epsilon_1}
      {p_1\cdot k_1} - \frac{p_2\cdot \epsilon_1} {p_2\cdot k_1}
       \right]\left[{\cal M}_{\gamma e\rightarrow Ze}\right]
\nonumber\\
&=& -\frac{\kappa}{2e^3}
     \frac{p_1\cdot k_1 p_2\cdot k_1}{k_1\cdot k_2}
     \left[{\cal M}_{\gamma s\rightarrow Ys}\right]
    \left[{\cal M}_{\gamma e\rightarrow Ze}\right],
    \label{tgeze}
\end{eqnarray}
where ${\cal M}_{\gamma e \rightarrow Ze}$ is the transition
amplitude of an electroweak process,
\begin{eqnarray}
&&{\cal M}_{\gamma e\rightarrow Ze}=
\frac{ieg_Z}{2}
   \bar{u}(p_2,s_2)
\left[\not\!{\epsilon^*_2}
     \frac{(\not\!{k_1}\not\!{\epsilon_1}+2p_1\cdot
       \epsilon_1)} {k_1\cdot p_1}
 - \frac{(2p_2\cdot \epsilon_1
       -\not\!{\epsilon_1} \not\!{k_1})}
    {k_1\cdot p_2}
   \not\!{\epsilon^*_2}\right]
\nonumber\\
&&\hskip 5cm
  \times
  \left[\epsilon_L(1-\gamma_5) +\epsilon_R(1+\gamma_5)\right]
      u(p_1,s_1).\label{reze}
\end{eqnarray}

Next, the transition amplitude ${\cal M}_{ge \rightarrow W\nu_e}$
is obtained from the Lagrangian ${\cal L}_{ge}$, ${\cal L}_{gW}$,
and ${\cal L}_{geW}$ as
\begin{eqnarray}
{\cal M}_{ge\rightarrow W\nu_e}
&=&{\cal M}^{W\nu_e}_a +{\cal M}^{W\nu_e}_b
  +{\cal M}^{W\nu_e}_c +{\cal M}^{W\nu_e}_d,
\end{eqnarray}
If $\epsilon_L$, $\epsilon_R$, and $g_Z$ in Eqs.~(\ref{mz1})
-(\ref{mz4}) are replaced by $1$, $0$, and $g_W$,
respectively, the transition amplitudes ${\cal M}^{W\nu_e}_i
(i=a,b,c,d)$ can be obtained in the same form as ${\cal M}^{Ze}_i$
 of the process
$ge\rightarrow Ze$.
Then the transition amplitude of the process $ge\rightarrow W\nu_e$
is completely factorized as
\begin{eqnarray}
{\cal M}_{ge\rightarrow W\nu_e}
&=& -\frac{\kappa}{2e}
     \frac{p_1\cdot k_1 k_2\cdot k_1}{p_2\cdot k_1}
     \left[\frac{p_1\cdot \epsilon_1} {p_1\cdot k_1}
      - \frac{k_2\cdot \epsilon_1} {k_2\cdot k_1}
       \right]
       \left[{\cal M}_{\gamma e\rightarrow W\nu_e}\right]
\nonumber\\
&=& \frac{\kappa}{2e}
    (p_1\cdot k_1)
     \left[\frac{p_1\cdot \epsilon_1} {p_1\cdot k_1}
      - \frac{p_2\cdot \epsilon_1} {p_2\cdot k_1}
       \right]
       \left[{\cal M}_{\gamma e\rightarrow W\nu_e}\right],
    \label{tgewn}
\end{eqnarray}
where ${\cal M}_{\gamma e\rightarrow W\nu_e}$ is the transition
amplitude of electroweak process defined as
\begin{eqnarray}
&&{\cal M}_{\gamma e\rightarrow W\nu_e}=
\frac{ieg_W}{2}
   \bar{u}(p_2,s_2)
\left\{\not\!{\epsilon^*_2}
        \frac{(\not\!{k_1} \not\!{\epsilon_1}
           +2p_1\cdot \epsilon_1)}
           {k_1\cdot p_1}
\right. \nonumber\\
&&\hskip 2cm \left.
 + \frac{2}{k_1\cdot k_2}
\left[(\epsilon_1 \cdot\epsilon^*_2)\not\!{k_1}
  -(\epsilon^*_2\cdot k_1)\not\!{\epsilon_1}
  -(\epsilon_1\cdot k_2)\not\!{\epsilon^*_2}
      \right]
      \right\}(1 - \gamma_5) u(p_1,s_1).
\label{rewn}
\end{eqnarray}

Using the simply factorized transition amplitudes
we can obtain the polarization effects of massive vector bosons.
Factorization allows us to use the well-known polarization
effects in the ordinary QED for the investigation of polarization
effects in the graviton processes.
The processes $ge\rightarrow Ze$ and $ge\rightarrow W\nu_e$
have the same polarization property as
$\gamma e\rightarrow Ze$ and $\gamma e \rightarrow W\nu_e$,
respectively \cite{JS},
because we have freedom to choose $\epsilon^\mu$ which makes
the first braket in Eq.~(\ref{tgeze}) is independent of
the graviton helicity \cite{Choi1,Choi2}.
The polarization of $Z^0$ and $W$ can be defined through the
density matrix \cite{SY}
\begin{eqnarray}
\rho^{\mu\nu}&=&\frac{1}{3} I^{\mu\nu}
-\frac{i}{2m}\epsilon^{\mu\nu\lambda\tau}k_{2\lambda}P_\tau
      -\frac{1}{2}Q^{\mu\nu}, \label{8a}\\
\end{eqnarray}
where
\begin{eqnarray}
I^{\mu\nu}=-\eta^{\mu\nu}+ \frac{k_2^\mu k_2^\nu}{m^2},
\end{eqnarray}
and $P_\tau$ and $Q^{\mu\nu}$ are called the polarization
vector and polarization tensor, respectively.
The explicit forms of the differential cross section and
the polarization vector and tensor of $Z^0$ in the
$ge\rightarrow Ze$ in the massless limit of the electron
are as follow;
\begin{eqnarray}
&&\left[\frac{d\sigma}{dt}\right]_{ge \rightarrow Z e}
 =\frac{\kappa^2 g_Z^2 t}{2^9 \pi u(s+u)^2 s^3}A_0,
	   \label{dsz}
\end{eqnarray}
\begin{eqnarray}
&&P^\mu=\frac{2}{m_Z}
   \left[i(|\epsilon_R|^2 +|\epsilon_L|^2)R_1^\mu
             + (|\epsilon_R|^2 -|\epsilon_L|^2)R_2^\mu
		\right]/A_0, \nonumber\\
&&Q^{\mu\nu}= -\frac{1}{3}I^{\mu\nu}
    - 2\left[(|\epsilon_R|^2 +|\epsilon_L|^2)R_1^{\mu\nu}
         + i(|\epsilon_R|^2 -|\epsilon_L|^2)R_2^{\mu\nu}
		\right]/A_0,
		\label{pqmu}
\end{eqnarray}
where $s$, $t$, and $u$ are the usual Mandelstam variables
and $A_0$, $R_1^\mu, R_2^\mu, R_1^{\mu\nu}$,
and $R_2^{\mu\nu}$ are defined \cite{JS} as
\begin{eqnarray}
&&A_0 =su\left\{-(|\epsilon_R|^2+|\epsilon_L|^2)
    \left[2t m_Z^2(1+\xi_3)+ s^2 +u^2 \right]
   +\xi_2(|\epsilon_R|^2-|\epsilon_L|^2) (s-u)(t+m_Z^2)
    \right\},
\nonumber\\
&&R_1^\mu= \frac{i}{2}su \xi_2
     \left[2m_Z^2(tk_1^\mu-sp_1^\mu+up_2^\mu)
     +(s^2+u^2)k_2^\mu\right],
           \nonumber\\
&&R_2^\mu=su m_Z^2\left\{ (s-u)K^\mu
        -(1+\xi_3)\left[A^\mu+t(p_1+p_2)^\mu\right]
        + 2\xi_1<\mu k_1 p_1 p_2>
          \right\} ,
          \nonumber\\
&&R_1^{\mu\nu}= \frac{su}{t}\left\{
         -t(su I^{\mu\nu}-2m_Z^2 K^\mu K^\nu)\right.
  \nonumber\\
    &&  \hskip 1cm -(1-\xi_3)
   \left[ t^2 k_1^\mu k_1^\nu +sut \eta^{\mu\nu}
      -tk_1^\mu(up_1^\nu-sp_2^\nu)
      -tk_1^\nu(up_1^\mu-sp_2^\mu)
      +A^\mu A^\nu\right]
\nonumber\\
 &&\hskip 1cm
  +(1+\xi_3)\left[A^\mu+t(p_1+p_2)^\mu\right]
              \left[A^\nu+t(p_1+p_2)^\nu\right]
\nonumber\\
  &&\hskip 1cm \left.  - 2\xi_1\left[<\mu k1 p1 p2>
              \left(A^\nu+t(p_1+p_2)^\nu\right)
               +(\mu \leftrightarrow \nu)\right]
\right\},
\nonumber\\
&&R_2^{\mu\nu}=-\frac{i}{2}su \xi_2
	  \left\{(u-s) (t+m_Z^2)
            \frac{k_2^\mu k_2^\nu}{m_Z^2}
     +2\left[m_Z^2k_1^\mu(p_1+p_2)^\nu-k_2^\mu A^\nu
         +(\mu \leftrightarrow \nu)\right]\right\}.
	  \label{qrmu}
\end{eqnarray}
In Eqs. (24) the $\xi_i$ $(i=1,2,3)$ are Stokes parameters
of the graviton beam \cite{Choi2} and $A^\mu$, $K^\mu$, and
$<\mu k_1 p_1 p_2>$ are defined as
\begin{eqnarray}
&&A^\mu=up_1^\mu + s p_2^\mu, \nonumber\\
&&K^\mu=k_1^\mu -\frac{k_1\cdot k_2}{m_Z^2}k_2^\mu
         =k_1^\mu -\frac{(s+u)}{2m_Z^2}k_2^\mu,
\nonumber\\
&&<\mu k_1 p_1 p_2>=
 \epsilon^{\mu\nu\alpha\beta}
 k_{1\nu} p_{1\alpha} p_{2\beta}.
\end{eqnarray}

The corresponding values for the $ge \rightarrow W \nu_e$
process can be obtained from Eqs.~(\ref{dsz})-(\ref{qrmu})
by replacing $\epsilon_L$, $\epsilon_R$, $m_Z$, and $g_Z$
by $1$, $0$, $m_W$, and $g_W$.
The effect of polarization of gravity can be obtained
through its effect to $P^\mu$ and $Q^{\mu\nu}$.
It is noted that the graviton can couple to any particle
and the two processes $ge \rightarrow Ze$ and
$ge \rightarrow W \nu_e$ have the four Feynman diagrams
of the same form as given in Fig.~1.
Among them we have chosen two independent diagrams in
order to compare them with the result of the electroweak
theory as in Eqs.~(\ref{reze}) and ~(\ref{rewn}).
However, in the standard model of the electroweak theory
two independent diagrams are different, but they are
related through a mathematical identity as shown in
Ref. \cite{JS}.
This is a special case of universality held in the SM of
electroweak theory.
The factorization is independent of the chirality of
$Z^0$, $W$ couplings to the fermion line and mass does
not affect the factorization at all.
\section*{Acknowledgments}

The work was supported in part by KOSEF (the Korea Science and
Engineering Foundation) through the SRC program and in part by
the Korean Ministry of Education. J. S. Shim would like to
thank Kyung Hee University for the postdoctoral fellowship.

\newpage
\section*{Figure Caption}
\begin{enumerate}
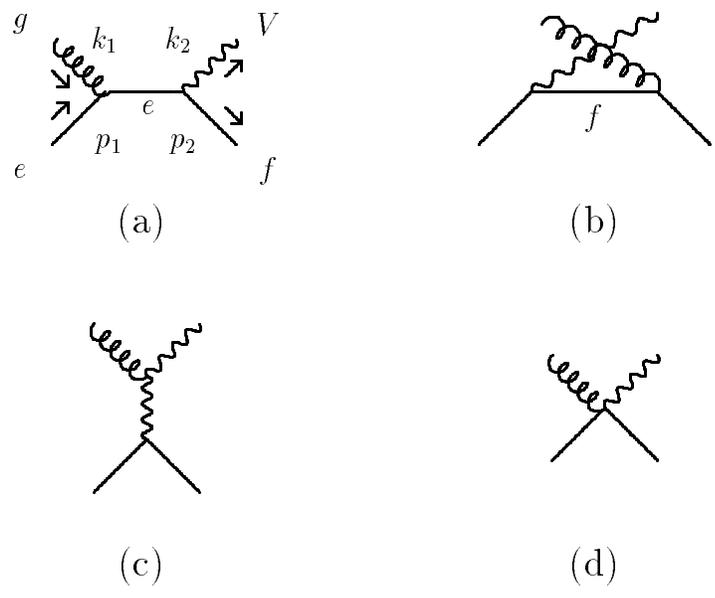
\item[{\bf Fig.~1}]
     Feynman diagrams for the process $ge\rightarrow Vf$.
     The curly line is for a graviton.
     $V$,  denoted  by a wiggly line, can be
     $Z$ or $W$.
     $f$,  denoted  by a solid line, can be
     a electron or a neutrino.
\end{enumerate}

\end{document}